\documentclass{jfm}

\usepackage{graphicx}
\usepackage{newtxtext}
\usepackage{newtxmath}
\usepackage{natbib}
\usepackage{hyperref}
\usepackage{amsmath}
\usepackage{xcolor}

\usepackage{soul}
\definecolor{lightblue}{rgb}{0.7,1.0,0.7}
\sethlcolor{lightblue}

\definecolor{lightred}{rgb}{1.0,0.7,0.7}
\definecolor{lightyellow}{rgb}{1.0,1.0,0.6}

\hypersetup{
    colorlinks = true,
    urlcolor   = blue,
    citecolor  = black,
}

\newcommand{\RomanNumeralCaps}[1]
\linenumbers

\title{Enhanced Classical Nucleation Theory for Cavitation Inception in the Presence of Gaseous Nuclei}

\author{Mazyar Dawoodian\aff{1}
  \corresp{\email{mazyar.dawoodian@uni-due.de}},
  Ould el Moctar\aff{1}}

\affiliation{\aff{1}Institute for Sustainable and Autonomous Maritime Systems, University of Duisburg-Essen, 47057 Duisburg, Germany}

\begin{document}
\maketitle

\begin{abstract}
This paper introduces an enhanced Classical Nucleation Theory model to predict the cavitation inception pressure and to describe the behavior of  nanoscale gaseous nuclei during cavitation. Validation is achieved through molecular dynamics simulations. The findings highlight the significant role of nanoscale gaseous nuclei in lowering the tensile strength required for cavitation initiation. The results show that our enhanced CNT model predicts lower cavitation pressures than the Blake threshold, closely matching molecular dynamics simulations. Finally, our results illustrate that differences between cavitation pressures using the Van der Waals and ideal gas models are greatest for smaller nuclei and lower temperatures.
\end{abstract}


\section{Introduction}
\label{sec:Introduction}
Water's strong cohesion enables it to withstand tension for prolonged durations \citep{Borkent2009, Brennen2013}. Experimental measurements shows that highly negative pressures can be sustained before bubble nucleation \citep{Azouzi2013, Green1990, Pallares2014, Zheng1991}. Recent research has focused on cavitation in water under tension due to its significance in various biological processes \citep{Ohl2006, Rowland2015, Vincent2012, Wheeler2008} and other industrial and medical applications \citep{Cui2018, Lohse2018, Sarc2018}, although its efficiency remains limited. Additionally, cavitation in water poses challenges in turbine and propeller design \citep{Kumar2010, Rezaee2024, Wu2018}. At high temperatures, different methods converge as the liquid cannot sustain significant tension. However, a much higher metastability level is achieved when examining cavitation in inclusions along an isochoric path \mbox{\citep{Azouzi2013, Green1990, Zheng1991}} compared to other techniques \mbox{\citep{Davitt2010, Herbert2006}} at low temperatures \mbox{\citep{Caupin2006}}. Critical cavitation pressure depends on factors like temperature, viscosity, surface tension, time, and strain rate \citep{Mahmud2020, Menzl2016}, with surface tension and temperature being most influential \citep{Fisher1948}.

Cavitation in practice often arises from pre-existing nuclei, a well-investigated phenomenon \citep{Caupin2006, Lohse2015, Gallo2020, sharma2024, Jones1999}. The pioneering work of \citet{Harvey1944} demonstrated the impact of pre-existing gas cavities on cavitation, attributing the measured low tensile strength on gaseous nuclei, a perspective subsequently supported by numerous studies. In purified and degassed liquids, with suppressed micrometre-sized nuclei, \citep{Herbert2006}, the measured tensile strengths (less than 30 MPa) remain notably lower than the theoretically expected 140 MPa \citep{Azouzi2013, Zheng1991}, leaving the cause of this discrepancy unclear. \citet{Morch2019} investigated the potential role of nanoscale gas bubbles/droplets as nuclei, which may be stabilized by the surface tension forces to achieve a gas diffusion balance with the adjacent liquid. \citet{Zhou2020} further explored the stabilization of nanobubbles on a heterogeneous substrate through molecular dynamics simulations. \mbox{\cite{hong2022}} have demonstrated that bulk nanobubbles act as non-condensable gaseous nuclei, significantly lowering the acoustic cavitation threshold in water. However, the effects of such gaseous nanoscale nuclei on the cavitation dynamics remain poorly understood due to the challenging in observing them. 

According to classical theories \mbox{\citep{Epstein1950}}, nanobubbles are expected to dissolve rapidly due to the high Laplace pressure \mbox{\citep{Alheshibri2016}}. However, recent advancements show that they can survive for much longer periods, ranging from hours to even months \mbox{\citep{zhou2021generation}}. For instance, \mbox{\cite{Chen_PRL2024}} demonstrated that thermal capillary waves lower surface tension, counterbalancing the Laplace pressure and, thus, preventing rapid dissolution. By performing molecular dynamics simulations, they confirmed the stability of bulk nanobubbles even after 50 ns. \mbox{\cite{Tan_PRL2020}} proposed electrostatic stabilization to predict equilibrium radii of 50–500 nm, which were consistent with experimental observations from dynamic light scattering and electron microscopy. \mbox{\cite{Nirmalkar2018}} reported bulk nanobubbles (100–120 nm) lasting for months, stabilized by surface charge effects. \mbox{\cite{jadhav2020}} demonstrated stability in aqueous organic solvents for over three months due to gas oversaturation. \mbox{\cite{jin2020}} tracked microbubble shrinkage into stable nanobubbles (280 nm, zeta potential -33 mV), using dark-field microscopy, while \mbox{\cite{hewage2022}} showed dynamic equilibrium under supersaturation. Additional support from \mbox{\cite{wang2022theoretical}}, \mbox{\cite{wang2019generation}}, \mbox{\cite{ke2019formation}}, and \mbox{\cite{rossello2023clean}} further validated the persistence of bulk nanobubbles, reinforcing their role as nucleation sites in our model. \mbox{\cite{gao2021-Langmuir}} employed molecular dynamics simulations to reveal that bulk nanobubble stabilization results from a combination of factors, including high internal gas density, electrostatic surface charges, and weakened interfacial hydrogen bonding. \mbox{\cite{Ghaani2020}} demonstrated that external electric fields can massively enhance bulk nanobubble formation and stability. \mbox{\cite{fang2020formation}} showed that vibration can induce the formation of stable bulk nanobubbles in water, with sizes typically depending on vibration frequency and duration. \mbox{\cite{li2024}} demonstrated that nanoscale bulk gaseous nuclei play a crucial role in laser-induced cavitation, significantly influencing bubble dynamics.

Classical nucleation theory (CNT) is a foundational framework widely used to study homogeneous cavitation, with its predictive reliability validated by experiments \citep{Azouzi2013} and molecular dynamics simulations {\citep{Menzl2016}. CNT forms the foundation for modern nucleation models, including density functional theory \mbox{\citep{Oxtoby1988}} and kinetic nucleation theory \mbox{\citep{Shen2003}}. These models have also been extended to heterogeneous cavitation, including cavitation on smooth, rigid surfaces \mbox{\citep{Blander1975}}. However, \citet{Menzl2016} revealed that CNT fails quantitatively at nanoscopic scales because it lacks critical microscopic information, including the curvature dependence of surface tension and thermal fluctuations that influence bubble expansion. \citet{Lohse2016} emphasized the significance of these findings, noting that Menzl et al.’s work successfully bridges macroscopic and nanoscopic descriptions. They suggested that integrating molecular dynamics with continuum mechanics could enhance our understanding of fluid mechanics at small scales and improve broader predictive models. Recent studies by \citet{Gao2021} and \cite{Alame2024} further illuminate the complexity of cavitation phenomena. \citet{Gao2021} extended CNT for nanoscale nuclei and showed that nanoscale nuclei significantly reduce the tensile strength. \cite{Alame2024}, proposed a Gibbs free energy framework linking homogeneous and heterogeneous nucleation, demonstrating gas content's role in stabilizing nuclei and lowering energy barriers.

We present an enhanced CNT model to predict the cavitation inception pressure and to describe the behavior of nanoscale gaseous nuclei during cavitation. We explicitly incorporate Van der Waals corrections to account for intermolecular forces within small gas nuclei and the Tolman length to account for curvature effects on surface tension  (Section \mbox{\ref{sec:theory}}). These refinements enable a more accurate assessment of cavitation induced by nanoscale gaseous nuclei. We validate the model by carrying out molecular dynamics simulations for nuclei a few nanometers in size (Section \ref{sec:valid2}). In Section \ref{sec:IG}, we further analyze deviations between Van der Waals and ideal gas predictions. In Section \mbox{\ref{sec:vsBlake}}, we compare cavitation inception pressures from our enhanced CNT model, from the Blake threshold, and from MD simulations across different temperatures and nucleus sizes to assess their accuracy. Finally, Section \ref{sec:Conclusions} presents the conclusions.

\section{Cavitation at Gaseous Nuclei – Theoretical Model}\label{sec:theory}
Numerous models have been developed to describe cavitation. The classical model, dating back to \citet{volmer1926}, focuses on cavitation in homogeneous solutions from a thermodynamic perspective. Extensive reviews on CNT are available in the books by, \citet{SKRIPOV1974}, and \citet{Carey2020}, as well as in articles by \citet{Blake1949tensile} and \citet{Bernath1952}. \citet{Harvey1944} introduced the concept of pre-existing gas cavities on surfaces as potential cavitation nuclei, stabilized by surface geometry and contact angles. Additional models consider micrometer-sized, skin-stabilized gas bubbles in liquid \citep{Yount1979} or attached to solid surfaces \citep{Andersen2015}. While these models provide valuable insights, they fall short in explaining the high tensile strength observed in highly purified, degassed water, which experimental measurements exceed compared to theoretical predictions. This discrepancy has led to speculation that it may be due to nanoscale nuclei, such as suspended nanoscale gas bubbles or droplets \citep{Morch2019}, which are challenging to eliminate.

\begin{figure}
  \centerline{\includegraphics[scale=0.28]{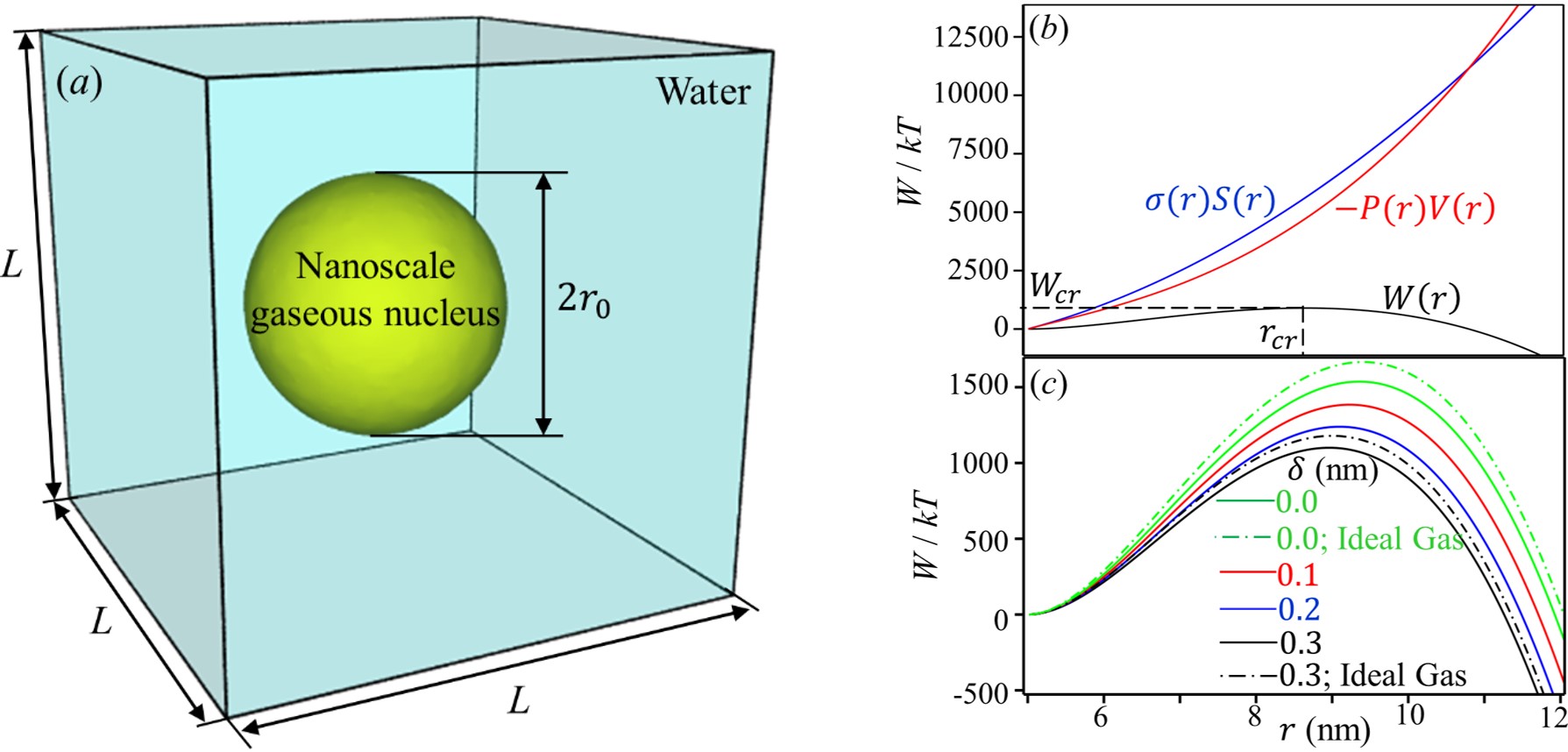}}%
  \caption{(a) A spherical nanoscale gaseous nucleus suspended in liquid. (b) Energy barrier for forming of a bubble of radius $r$ from a nanoscale gaseous nucleus (black line). The blue and red lines are the first and second terms in eq~\eqref{eq3}, here $r_0=5$ nm, $\delta=0.3$ nm, $P_l=-15$ MPa, and $T=298$ K. (c) This plot shows the work required to expand a gas pocket ($r_0=5$ nm), for various $\delta$, and reflects the balance between the free energy cost of increasing the liquid–gas interface, which dominates for smaller bubbles, and the mechanical work gained from expansion under tension, favoring larger bubbles. The peaks represent the critical radius, where the free energy barrier is at its maximum.}
\label{fig1}
\end{figure}

In this study, nanoscale gaseous nuclei are modeled with specific attention to their physical properties, rather than assuming them as simple equivalent voids. Unlike a pure void, a gaseous nucleus exhibits distinct characteristics, such as internal gas pressure, surface tension, and molecular interactions with the surrounding liquid. These factors significantly impact the cavitation process, especially under high metastability conditions. Our model incorporates the actual gas content and nanoscale effects to provide a more accurate representation of cavitation behavior. The nuclei are assumed to be uniformly distributed in the liquid with an initial radius $r_0$, and the surrounding liquid is treated as far from any container walls to avoid boundary effects. Under these conditions, cavitation is expected to initiate at the gaseous nuclei, and with this refined model we aim to capture the influence of the gaseous nature of the nuclei on cavitation inception. The following analysis documents the derivation of a mathematical model to predict cavitation pressure variations based on the properties of these nuclei and the liquid.

In the framework of CNT, cavitation is an activated process in which a free energy barrier must be surpassed to form a critical nucleus, after which the bubble can grow spontaneously. The work $W$ required to form a bubble with radius $r$ from a nanoscale gas pocket with radius $r_0$ (figure \ref{fig1}a) is expressed as follows:

\begin{equation}
  W(r,P_g)=4\pi(r^2-r_0^2)(1-\frac{2\delta}{r})\sigma_0+\frac{4\pi}{3}(r^3-r_0^3)(P_l-P_g-P_v)
  \label{eq1}
\end{equation}

where ~$\sigma_0$ is the surface tension of the planar interface, $P_l$, $P_g$, and $P_v$ represent the liquid, non-condensable gas, and vapor pressures, respectively. \eqref{eq1} incorporates the curvature dependence of the surface tension through a Tolman-like correction, where Tolman length $\delta$ accounts for changes in surface tension as interface curvature increases, a key factor in nanoscale bubbles \citep{Menzl2016}. The first term in the RHS of \eqref{eq1} represents the energy cost required to expand the liquid–gas interface area, while the second term of \eqref{eq1} corresponds to the energy gained from the increase in the bubble volume. In this study, the vapor pressure is taken to be the saturated vapor pressure $P^{s}_{v}$ \citep{Hao1999}. To estimate $P_g$, we employ the Van der Waals equation to account for non-ideal gas behavior at the nanoscale, which incorporates the effects of intermolecular attractions and the finite volume of gas molecules that are critical in small gas clusters. These corrections ensures more accurate modeling under conditions of high pressure and small volumes, where the assumptions of the ideal gas law no longer hold. The Van der Waals equation is expressed as:

\begin{equation}
  P_g=\frac{nRT}{\frac{4}{3}\pi r^3-nb}-\frac{9n^2 a}{16 \pi^2 r^6}
  \label{eq2}
\end{equation}

where $n$ is the moles of gas in the bubble, $R$ the universal gas constant, $T$ the temperature, $a$ and $b$ are the Van der Waals constants for intermolecular attraction and finite molecular volume, respectively. Combining \eqref{eq1} and \eqref{eq2}, the energy barrier that must be overcome for its formation is expressed as follows:

\begin{equation}
  W(r)=\sigma(r)S(r)+P(r)V(r)
  \label{eq3}
\end{equation}

where 

\[ \sigma(r)=(1-2\delta/r)\sigma_0 \text{ ,} \]
\[ S(r)=4\pi (r^2-r_0^2) \text{ ,} \]
\[ P(r)=P_l-\frac{nRT}{\frac{4}{3}\pi r^3 -nb}+\frac{9n^2 a}{16 \pi^2 r^6}-P_v^s \text{ , and} \] 
\[ V(r)=\frac{4\pi}{3}(r^3-r_0^3) \text{ .} \]

Figure~\ref{fig1}(b) shows the contributions of the terms of equation \eqref{eq3}. The surface energy term, $\sigma(r)S(r)$, increases with $r$ due to the growing surface area, while the volume term, $-P(r)V(r)$, initially increases the total work, but ultimately it counterbalances the surface energy at larger $r$. This interplay results in an energy barrier that peaks at the critical radius $r_{cr}$, beyond which bubble growth becomes spontaneous. In Fig. \ref{fig1}c, the work required to expand the nucleus is shown for various Tolman lengths ($\delta$) ranging from 0 to 0.3 nm \citep{Wilhelmsen2015}. As $\delta$ increases, both the critical work required to overcome the energy barrier and the critical radius decrease. This emphasizes the importance of incorporating curvature-dependent surface tension in nanoscale cavitation models. In this work, we select $\delta=0.3$ nm \citep{Magaletti2021, Wilhelmsen2015}. The plot also highlights the difference between the Van der Waals and the ideal gas models in predicting the work necessary for bubble expansion. The relationship between the critical radius and the liquid pressure can be derived through $\frac{dW(r)}{dr}=0$. After some simplification, we obtain the following equation:

\begin{equation}
\begin{split}
  P_l=P_v^s-2\frac{\sigma_0}{r_{cr}} \left( 1-\frac{\delta}{r_{cr}} \left( 1+\frac{r_0^2}{r_{cr}^2} \right) \right) + \left( -\frac{4nRT{\pi}r_{cr}^3}{3 \left( \frac{4 \pi r_{cr}^3}{3}-nb  \right)^2 } +  \frac{9 n^2 a}{8\pi^2 r_{cr}^6} \right) (1-\frac{r_0^3}{r_{cr}^3}) + \\
  \frac{nRT}{\frac{4 \pi r_{cr}^3}{3}-nb} - \frac{9n^2 a}{16 \pi^2 r_{cr}^3}
  \label{eq4}
\end{split}
\end{equation}

\begin{figure}
  \centerline{\includegraphics[scale=0.27]{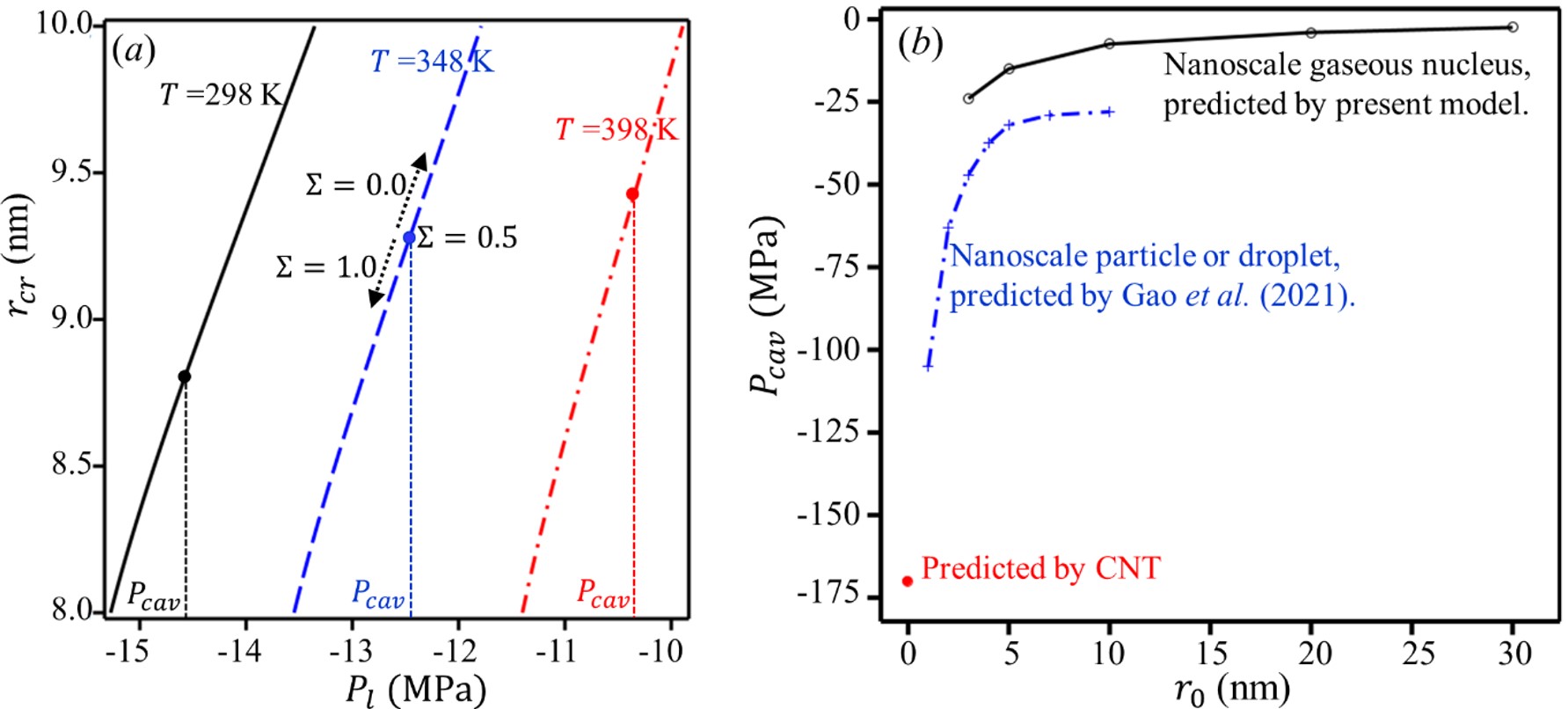}}%
  \caption{(a) This curve represents equation \eqref{eq4}, illustrating the relationship between cavitation probability $\Sigma$, $P_l$, $T$ and $r_{cr}$, for $r_0=5$ nm. The cavitation pressure $P_{cav}$ is defined as the liquid pressure at which $\Sigma=0.5$. As $T$ increases, the magnitude of the $P_{cav}$ required for cavitation decreases. (b) Comparison of water cavitation pressures as predicted by the present model, Gao's model, and CNT. This plot also demonstrates the influence of nuclei size on cavitation pressure.}
\label{fig2}
\end{figure}

Equation \mbox{\eqref{eq4}} involves the critical radius \mbox{\(r_{cr}\)} and the liquid pressure \mbox{\(P_l\)} (figure \mbox{\ref{fig2}}a). By substituting the expression for \mbox{\(r_{cr}\)} in \mbox{\eqref{eq3}}, the energy barrier of a critical nucleus can be expressed as follows:

\begin{equation}
  W_{cr}=\sigma(r_{cr})S(r_{cr})+P(r_{cr})V(r_{cr})
  \label{eq5}
\end{equation}

and the equation predicting the rate of cavitation at nanoscale nuclei $J$, can be expressed as:

\begin{equation}
  J=J_0 exp \left[ -\frac{W_{cr}}{kT} \right]
  \label{eq6}
\end{equation}

where $J_0=4 \sqrt{2 \sigma / \pi m} r_0^2 N_0^{2/3}/V$, $N_0$ is the molecular number density, and $V$ is the volume of the object in figure \ref{fig1}a. Assuming that the liquid pressure is maintained at $P_l$ and the tensile stress is applied for a time $\Delta t$, the probability of cavitation $\Sigma$ is given by (figure \ref{fig2}a):

\begin{equation}
  \Sigma=1-exp(-JV \Delta t)
  \label{eq7}
\end{equation}

In this study, we adopt the definition of cavitation pressure $P_{cav}$ as outlined by \citet{Caupin2012} where $P_{cav}$ is the liquid pressure $P_l$ at which $\Sigma = 1/2$. Then, by applying equations \eqref{eq5} and \eqref{eq6}, the value of $r_{cr}$ is determined. This leaves $P_l$ as the only unknown in equation \eqref{eq4}, representing our $P_{cav}$. Figure \ref{fig2}a shows, as the temperature increases, the liquid pressure required for cavitation decreases, and the critical radius becomes larger. The vertical lines indicate the specific cavitation pressures at each temperature, corresponding to the point where $\Sigma=1/2$. As seen, the probability of cavitation increases as the $P_l$ decreases. Recall that in CNT, the tensile strength of a liquid depends on both the waiting time $ \Delta t$ and the liquid volume $V$ under tension \mbox{\citep{Brennen2013}}. From equations \eqref{eq6} and \eqref{eq7}, the relation between $V \Delta t$ and $P_{cav}$ shows that $P_{cav}$ is only weakly affected by $V \Delta t$ as it appears in a logarithmic term. For example, \citet{Zheng1991} demonstrated that the predicted tensile strength changes by less than $5 \%$ when $\Delta t$ varies between $0.001 s$ and $1000 s$, with $V$ held constant. For an ideal gas $(a=b=0)$, the following equation is obtained:

\begin{equation}
  P_l=P_v^s-2 \frac{\sigma_0}{r_{cr}} \left( 1-\frac{\delta}{r_{cr}} \left( 1+\frac{r_0^2}{r_{cr}^2} \right) \right) + \frac{3nRT}{4 \pi r_{cr}^3} \frac{r_0^3}{r_{cr}^3}
  \label{eq8}
\end{equation}

which for uncorrected surface tension, $\delta=0$, is reduced to:

\begin{equation}
  P_l=P_v^s-2 \frac{\sigma_0}{r_{cr}} + \frac{3nRT}{4 \pi r_{cr}^3} \frac{r_0^3}{r_{cr}^3}
  \label{eq9}
\end{equation}

Following \citet{Herbert2006} and \citet{Gao2021}, we calculate the cavitation pressure with parameters $V = 2.1 \times 10^{-4} mm^3 $, and $\Delta t = 4.5 \times 10^{-8} s$.  In this study, oxygen (\( \text{O}_2 \)) is used as the non-condensable gas, with Van der Waals constants \( a = 1.382 \times 10^{-1} \, \text{m}^6 \cdot \text{Pa} / \text{mol}^2 \), \( b = 3.186 \times 10^{-5} \, \text{m}^3 / \text{mol} \). 

In figure \mbox{\ref{fig2}}b, the cavitation pressures on gaseous nucleus (as predicted by the present model) are compared to those for nanoscale particle or droplet (as predicted by \mbox{\citet{Gao2021}} model), along with classical nucleation theory. The present model predicts higher cavitation pressures for gaseous nuclei, whereas considering a nanoscale particle or droplet results in a lower cavitation pressure. The CNT prediction, represented as a single point, corresponds to an idealized case without nanoscale effects, leading to a much lower cavitation pressure. This comparison highlights the necessity of incorporating gaseous effects in cavitation modeling, as they significantly influence the predicted inception pressures.

\section{Comparison with Molecular Dynamics (MD) Simulations}\label{sec:valid2}
To validate the modified CNT model at smaller scales, we compare its predictions with MD simulations for nuclei with radii of $3–5$ $nm$. The MD solver GROMACS \citep{Hess2008} is used with the TIP4P/2005 water molecular model \citep{Abascal2005}, known for accurately replicating water properties, including surface tension \citep{Alejandre2010}. Simulations are conducted in a $30 \times 30 \times 15 nm^3$ water box containing 600,000 water molecules, with 3D periodic boundary conditions. Temperature control is achieved through a Nose-Hoover thermostat. The system was equilibrated under an NPT ensemble at 0.1 MPa and 298 K for 1 ns. The gaseous nucleus was assumed to consists only oxygen (O\textsubscript{2}) molecules, and their number is calculated using \eqref{eq2} (Van der Waals law). In the equilibrium condition, considering the Laplace pressure inside bubble, we obtain the following expression:  

\begin{equation}
  P_{l,0} + \frac{2}{r_{0}} \left( 1-\frac{2\delta}{r_0} \right) \sigma_0 = \frac{nRT}{\frac{4}{3} \pi r_{0}^3 - nb} - \frac{9 n^2 a}{16 \pi^2 r_0^6}
  \label{eqMD}
\end{equation}

\begin{figure}[H]
  \centerline{\includegraphics[scale=0.29]{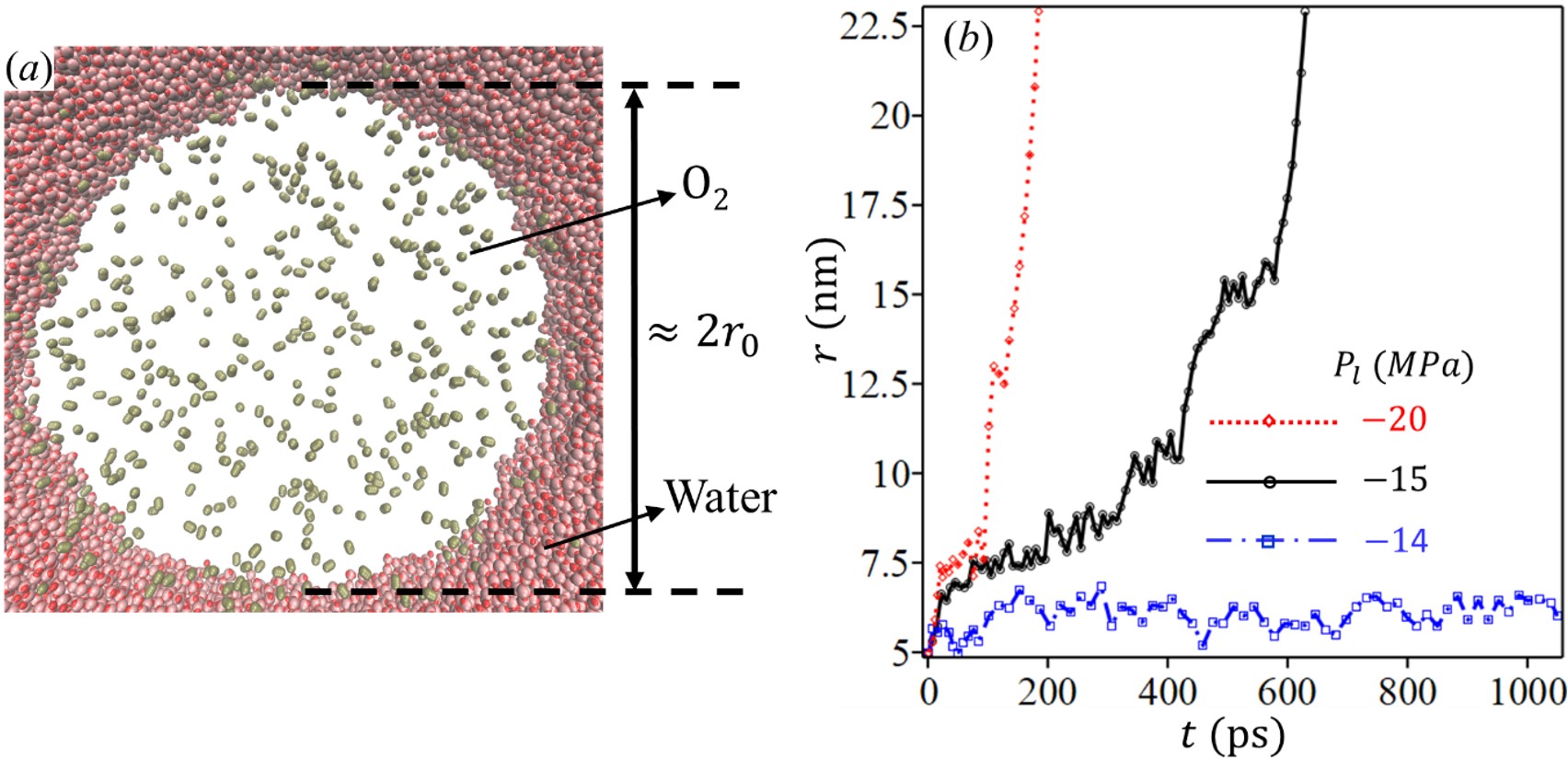}}%
  \caption{(a) Snapshot of the gaseous nucleus (central cross-sectional view), here $P_l=0.1$ MPa and $T=298$. (b) Time evolution of the nanobubble radius under different $P_l$. The curve corresponding to $P_l=-14$ MPa indicates no cavitation occurred. For lower pressures, cavitation did occur. The rate of nanobubble radius expansion increased at smaller $P_l$, here $r_0 \approx5$ nm and $T = 298$ K.}
\label{fig4}
\end{figure}

where $P_{l,0}=0.1$ MPa, and $\delta=0.3$ nm. In \eqref{eqMD}, the only unknown is number of gas molecules $n$ (Table \ref{tab:md}). These $n$ molecules of O\textsubscript{2} are then added to the box. 
To form a gaseous nucleus, we applied volume-controlled stretching, reducing system pressure to -180 MPa and creating a void that O\textsubscript{2} molecules rapidly fill due to a strong diffusion gradient. 

\begin{table}
  \begin{center}
\def~{\hphantom{0}}
  \begin{tabular}{cccccc}
       T [K]         & $298$    &   $323$  & $348$  & $398$  & $448$ \\[4pt]
       $r_0 = 3 nm$   & 1001 & 888  & 776  & 575  & 408 \\
       $r_0 = 5 nm$   & 3492 & 2988 & 2539 & 1816 & 1269\\
  \end{tabular}
  \caption{Number of O\textsubscript{2} molecules, $n$, inside the gaseous nucleus under the equilibrium condition, calculated using equation \mbox{\eqref{eqMD}}}.
  \label{tab:md}
  \end{center}
\end{table}

Within 0.5 ns, the void became densely populated, ensuring that (almost) all O\mbox{\textsubscript{2}} molecules diffuse into the cavity. The pressure is then gradually adjusted to 0.1 MPa, and the system equilibrates for 4 ns to allow the nucleus to stabilize (Fig. \mbox{\ref{fig4}}a). This method allows precise control over the nucleus diameter by adjusting the number of O\mbox{\textsubscript{2}} molecules, thereby improving reproducibility in our simulations. It is important to note that once the gaseous nucleus was formed, the surrounding liquid was no longer supersaturated, as most of the O\mbox{\textsubscript{2}} molecules had diffused into the void, leaving a negligible concentration of dissolved gas in the liquid phase.

\textcolor{blue}{
\subsection{Stability of the bulk nanobubble}}
\textcolor{blue}{Thermal capillary waves (TCWs) are thermal fluctuations occurring at gas-liquid interfaces \citep{Rowlinson2002}, arising from a balance between molecular thermal motion and intermolecular cohesion (surface tension), while gravitational and hydrodynamic effects remain negligible at this scale (generally below 1 nm). TCWs effectively explain the reduction in surface tension \citep{Chen_PRL2024}, as confirmed by direct optical observations \citep{Aarts2004, Derks-PRL-2006} and molecular dynamics simulations \citep{Zhao-POF-2023, Liu-PhysRevE-2023, Delgado-PhysRevLett-2008}. Reduced surface tension significantly contributes to nanobubble stability \citep{Chen_PRL2024}. Having previously clarified phenomena such as nanodroplet coalescence \citep{Perumanath-PhysRevLett-2019} and nanojet rupture \citep{Moseler-Sci-2000, Eggers-PhysRevLett-2002}, TCWs represent a critical interfacial property deserving thorough investigation within the nanobubble context.}

\textcolor{blue}{\cite{Chen_PRL2024}, in their recent notable publication, introduced a TCW-based Epstein–Plesset model demonstrating the long-term stability of bulk nanobubbles. Their findings indicate that a bulk nanobubble with an initial radius of 2.64 nm experiences only a modest 5\%  reduction in size over 100 ns, a result strongly corroborated by their direct molecular dynamics simulations. Importantly, TCWs arise naturally in molecular dynamics simulations employing the TIP4P/2005 potential, without artificial additions or modifications, as recently demonstrated explicitly by \cite{Chen_PRL2024}.}

\textcolor{blue}{In our MD model of a nanobubble with an initial radius of $r_0=3$ nm, we observed a slight reduction in bubble radius (about 4 \%) within the first 10 ns. In the subsequent 90 ns, the bubble radius remained nearly constant, resulting in only about a 7 \% total radius reduction over the entire 100 ns simulation. These findings align closely with the observations and model presented by \cite{Chen_PRL2024}. While the plain Epstein–Plesset theory (without incorporating TCW effects) predicts complete dissolution of the nanobubble (having similar size of $r_0=3$) within a much smaller time scale, the observed stability in our MD simulations confirms the significant role that TCWs play in sustaining nanobubble longevity.}

\subsection{Cavitation inception pressure}

\begin{figure}
  \centerline{\includegraphics[scale=0.4]{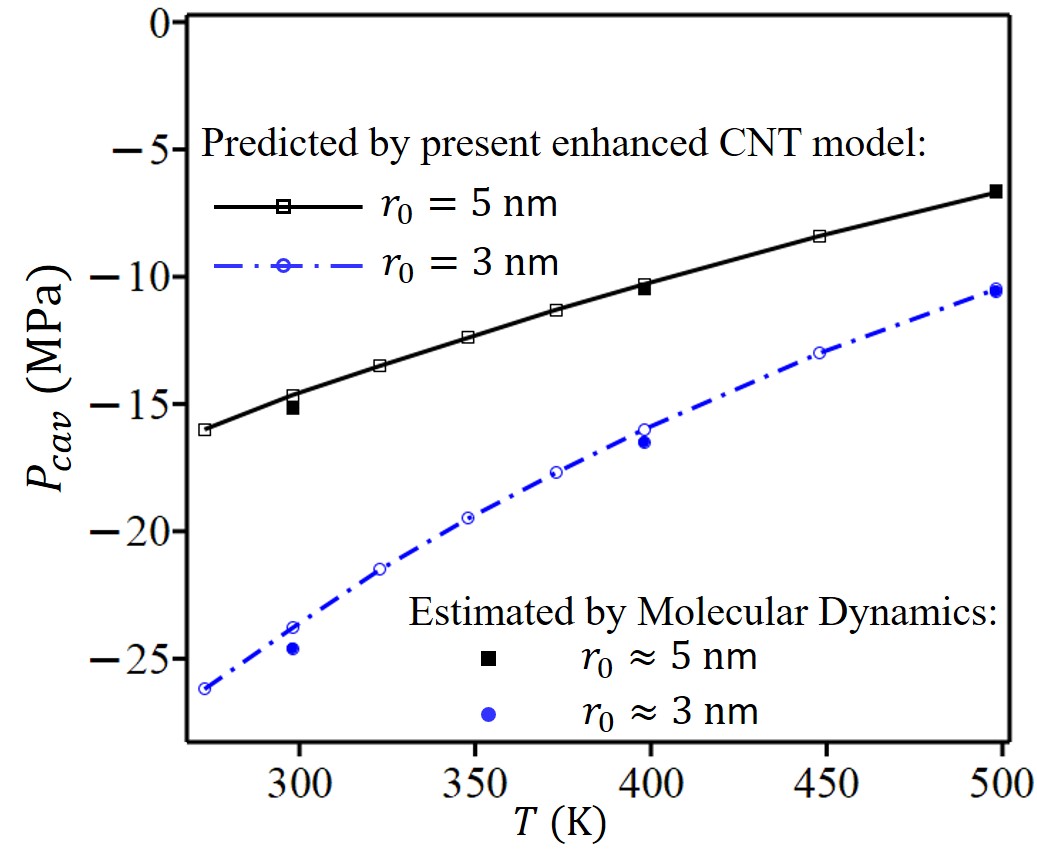}}%
  \caption{Comparison of $P_{cav}$ predicted by the proposed model and MD simulations for different $r_0$ and $T$.}
\label{fig-CNT-vs-MD}
\end{figure}

After stabilization, the pressure is reduced once again to initiate cavitation, analyzing system's response, now with a stable gaseous nucleus, under different negative pressures. Figure \ref{fig4}b shows the dynamic behavior of the gaseous nucleus radius under varying negative pressures. Starting with a radius about 5 nm, the nucleus remained stable at -14 MPa, indicating internal pressure balancing external pressure. At -15 MPa, cavitation occurs, causing rapid expansion, which accelerated further at pressures below -20 MPa. Figure \ref{fig-CNT-vs-MD} demonstrates that the cavitation pressure $P_{cav}$ predicted by the proposed mathematical model compares favorably to the MD simulation results for various equivalent nucleus sizes and liquid temperatures. The figure reveals a reduction in the liquid's tensile strength as the equivalent nucleus radius increases from 3 nm to 5 nm. Similarly, the tensile strength decreases as the liquid temperature rises from 298 K to 500 K. 

\section{Deviation from Ideal Gas Behavior}\label{sec:IG}
Figure~\ref{figIDgas} illustrates the deviation of the cavitation pressure predicted by the Van der Waals model ($P_{cav}$) from that predicted by the ideal gas law ($P_{cav}^{IG}$). The ratio (\( P_{\text{cav}} / P_{\text{cav}}^{IG} \)) shows at smaller nucleus radii, the deviation from the ideal gas prediction is more pronounced, especially at lower temperatures. This behavior is attributed to higher Laplace pressure in smaller nuclei, which amplified the effect of intermolecular forces and finite molecular size, as accounted for in the Van der Waals model. Conversely, as the nucleus radius increases, the predictions from both models converge, reflecting a diminished impact of real gas behavior. Additionally, the deviation decreases with rising temperature, as higher thermal energy reduces the influence of intermolecular forces, making the gas behave more ideally. The annotations in figure \mbox{\ref{figIDgas}} highlight this deviation by presenting specific cavitation pressures under different conditions. For instance, at $T=273$ K, the cavitation pressure reaches -29.3 MPa for the Van der Waals model and -23.8 MPa for the ideal gas model, while molecular dynamics simulations predict -24.6 MPa, thereby aligning more closely with the Van der Waals model.

\begin{figure}
  \centerline{\includegraphics[scale=0.4]{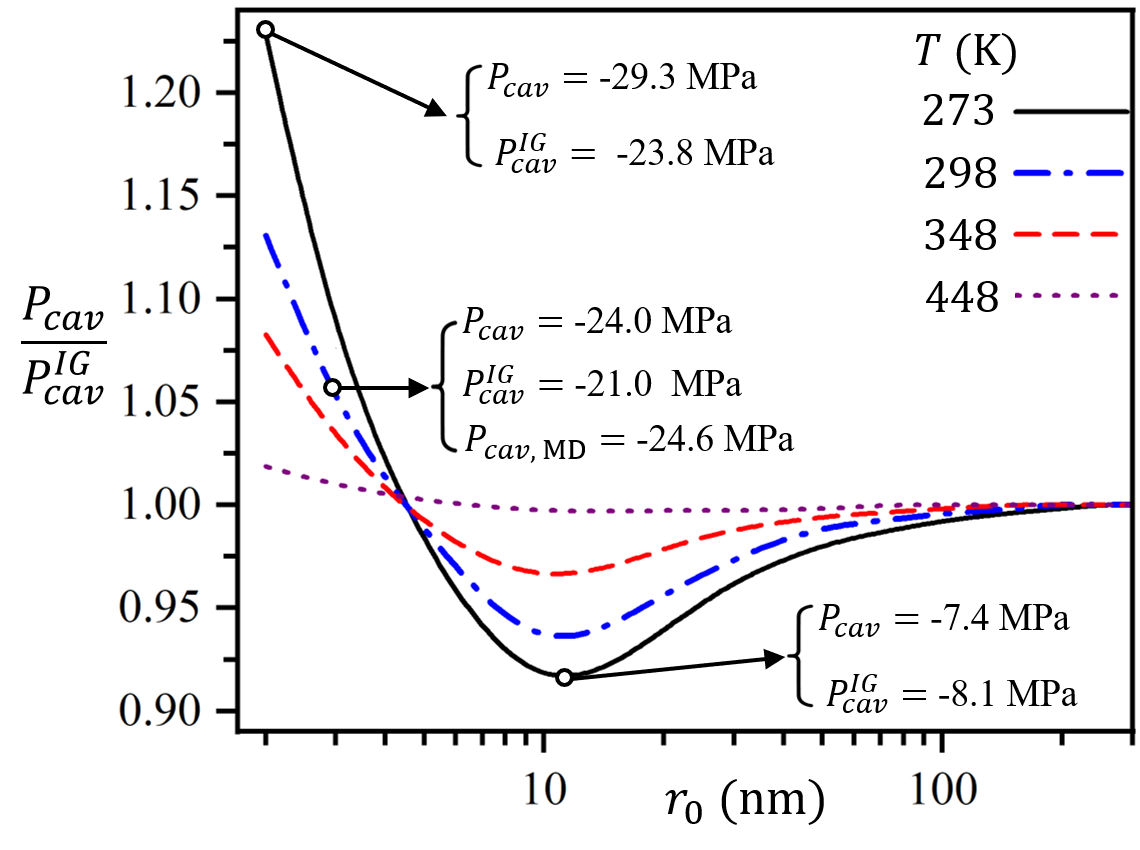}}%
  \caption{Deviation of cavitation pressure predicted by the Van der Waals model ($P_{cav}$) from that predicted by the ideal gas law ($P_{cav}^{IG}$). Deviations are higher for smaller $r_0$ and lower $T$. The annotations highlight specific cavitation pressures under different conditions. As temperature increases, the deviations between models decrease, indicating a reduced influence of intermolecular forces.}
\label{figIDgas}
\end{figure}

\section{Present model vs. Blake threshold}\label{sec:vsBlake}
In this section, we present a comparative analysis of cavitation inception pressures as predicted by the Blake threshold and our enhanced CNT. The Blake threshold \mbox{\citep{blake1949onset, leighton2012acoustic, neppiras1951, atchley1989}}, provides a criterion for the stability of gas bubbles in liquids. However, we have observed that the Blake threshold may not fully capture the complexities of cavitation phenomena, especially at nanoscale. Our enhanced CNT model incorporates additional factors to address these limitations, aiming to provide more accurate predictions of cavitation pressures. While previous studies have highlighted the limitations of the Blake threshold in predicting cavitation inception pressures for surface nanobubbles \mbox{\citep{borkent2007superstability, dockar2018mechanical}}, there is a lack of literature addressing its accuracy in predicting cavitation inception pressure on bulk nanobubbles. We review the Blake threshold formulas for Van der Waals gases, incorporating the Tolman length correction. Assuming a constant vapor pressure \mbox{\( P_v^s \)}, the ambient liquid pressure \mbox{\( P_{l} \)} is approximated as: 

\begin{equation}
  P_{l}=P_v^s-2\frac{\sigma_0}{r} \left( 1-\frac{2\delta}{r}  \right)  +  \frac{nRT}{\frac{4 \pi r^3}{3}-nb} - \frac{9n^2 a}{16 \pi^2 r^3}
  \label{eqBlake1}
\end{equation}

The critical bubble radius, denoted as \mbox{\( r_{cr,Blake} \)} can be derived through \mbox{$\frac{dP_{l}}{dr}=0$}. Upon finding \mbox{\( r_{cr,Blake} \)}, cavitation inception pressure based on Blake threshold \mbox{\( P_{cav,Blake} \)} will be expressed as:

\begin{figure}
  \centerline{\includegraphics[scale=0.455]{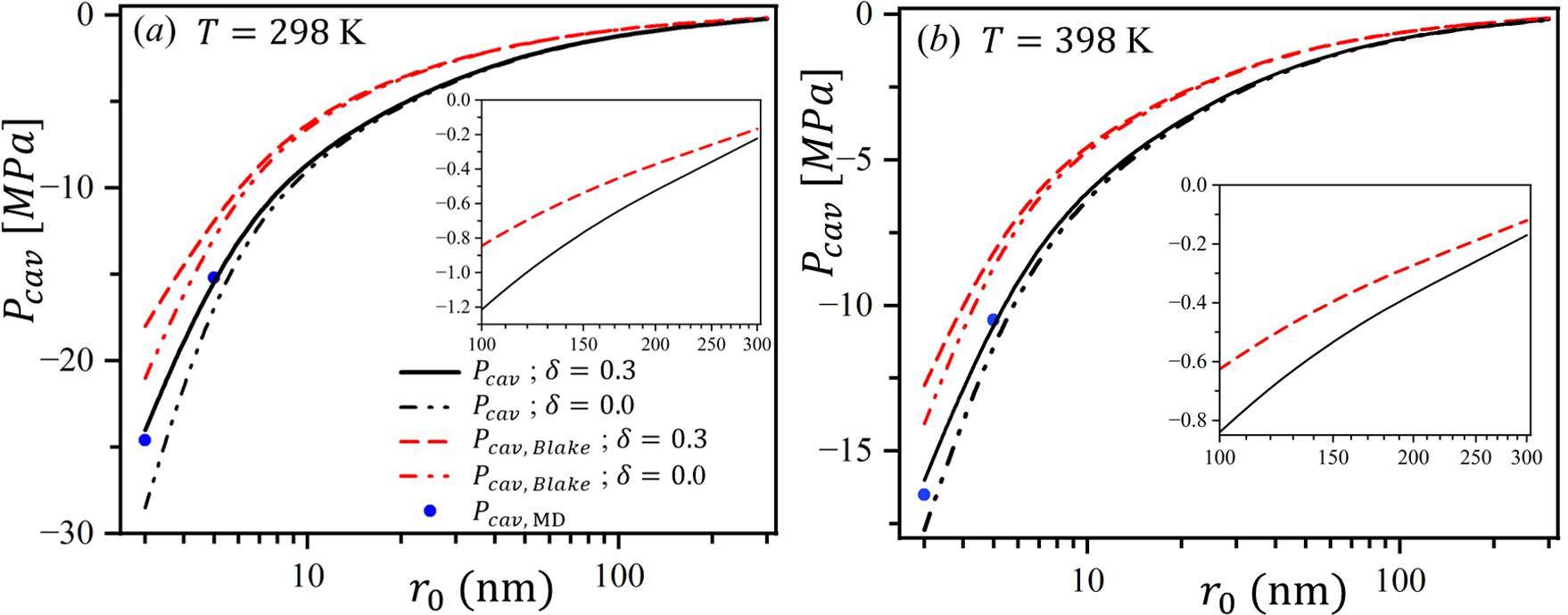}}%
  \caption{Comparison of \mbox{$P_{cav}$} predicted by the proposed model, the Blake threshold and molecular dynamics simulations for different \mbox{$r_0$}, at (a) \mbox{$T=298$} K and at (b) \mbox{$T=398$} K. The present enhanced CNT model predicts lower cavitation pressures than the Blake threshold and aligns more closely with molecular dynamics simulations.}
\label{fig-vs-Blake}
\end{figure}

\begin{equation}
  P_{cav,Blake}=P_v^s-2\frac{\sigma_0}{r_{cr,Blake}} \left( 1-\frac{2\delta}{r_{cr,Blake}}  \right)  +  \frac{nRT}{\frac{4 \pi r_{cr,Blake}^3}{3}-nb} - \frac{9n^2 a}{16 \pi^2 r_{cr,Blake}^3}
  \label{eqBlake2}
\end{equation}

Although  \mbox{$r_0$}  does not explicitly appear in the Eq. \mbox{\eqref{eqBlake2}}, it is inherently linked to  \mbox{$n$} through Eq. \mbox{\eqref{eqMD}}. We compare cavitation inception pressures from our enhanced CNT \mbox{(\( P_{cav} \))}, from the Blake threshold \mbox{(\( P_{cav,Blake} \))}, and from MD simulations \mbox{(\( P_{cav, \text{ MD}} \))} at 298 K and 398 K, and for different Tolman lengths \mbox{\( \delta \)} as a function of bubble radius, as shown in Fig. \mbox{\ref{fig-vs-Blake}}. Calculations show that the Blake threshold predicts higher cavitation pressures. For example, at \mbox{\( r_0 = 5 \)} nm and \mbox{\( T = 298 \)} K, it gives \mbox{\( P_{cav,Blake}=-11.6 \)} MPa, while the present model predicts \mbox{\( P_{cav}=-14.65 \)} MPa, highlighting a notable discrepancy between the models. While \mbox{\( P_{cav,Blake} \)} predicts higher cavitation pressures, the present enhanced CNT model (\mbox{\( P_{cav} \)}) closely matches MD results, at least for \mbox{\( r_0 \leq 5 \)} nm, where simulations were computationally feasible. This agreement indicates that our enhanced Classical Nucleation Theory provides a more accurate prediction than the Blake threshold within the studied range. The enhanced CNT model \mbox{\eqref{eq4}} achieves greater accuracy by incorporating higher-order corrections compared to the Blake threshold \mbox{\eqref{eqBlake2}}, enabling a more precise representation of nanoscale effects.

\section{Conclusions}\label{sec:Conclusions}
Experimental observations indicated that the water's tensile strength is consistently lower than theoretical estimates, even after purification and degassing, due to the unavoidable presence of nanoscale gaseous nuclei. This study presented an enhanced Classical Nucleation Theory model that integrated Van der Waals corrections to account for intermolecular forces and the Tolman length to adjust surface tension for curvature effects. The refined framework provided a cohesive description of cavitation inception in nanoscopic scales. Model validation was conducted through molecular dynamics simulations. These results highlighted the critical influence of nanoscale gaseous nuclei in reducing the water's tensile strength, offering a comprehensive tool for evaluating their role in cavitation phenomena. We also showed that the present enhanced CNT model provided more accurate cavitation inception pressure predictions than the Blake threshold, thereby closely matching molecular dynamics simulation results. Furthermore, the deviations between $P_{cav}$ predicted by Van der Waals and ideal gas models showed the importance of real gas behavior, particularly for small nuclei and low temperatures.

The discrepancy between \mbox{\( P_{cav} \)} predicted by present enhanced CNT model and Blake threshold for bulk nanobubbles still requires further investigation through experiments or molecular dynamics simulations. A significant challenge is the lack of experimental data on cavitation inception pressure in bulk nanobubbles, as their small size makes direct observation technically difficult. Similarly, the absence of molecular dynamics simulations addressing cavitation inception pressure in bulk nanobubbles represents a gap that future research should aim to fill.

\vspace{1cm}

\backsection[Acknowledgements]{The authors acknowledge the computing time granted by the Center for Computational Sciences and Simulation (CCSS) of the University of Duisburg–Essen, Germany on its supercomputer magnitUDE (DFG grants INST 20876/209-1 FUGG, INST 20876/243-1 FUGG).}

\backsection[Funding]{This research received no specific grant from any funding agency, commercial or not-for-profit sectors.}

\backsection[Declaration of interests]{The authors report no conflict of interest.}

\backsection[Author ORCIDs]{

Mazyar Dawoodian, https://orcid.org/0000-0002-4818-6225;

Ould el Moctar, https://orcid.org/0000-0002-5096-4436.}

\bibliography{library}

\end{document}